\documentclass[letterpaper,english,aps,prl,lengthcheck,superscriptaddress,showpacs,floatfix,amsfonts,amsmath,amssymb]{revtex4-1}
\usepackage[T1]{fontenc}
\usepackage[latin9]{inputenc}
\usepackage{amstext}
\usepackage{graphicx}
\usepackage{amssymb}

\makeatletter
\@ifundefined{textcolor}{}
{%
 \definecolor{BLACK}{gray}{0}
 \definecolor{WHITE}{gray}{1}
 \definecolor{RED}{rgb}{1,0,0}
 \definecolor{GREEN}{rgb}{0,1,0}
 \definecolor{BLUE}{rgb}{0,0,1}
 \definecolor{CYAN}{cmyk}{1,0,0,0}
 \definecolor{MAGENTA}{cmyk}{0,1,0,0}
 \definecolor{YELLOW}{cmyk}{0,0,1,0}
 }

\usepackage{bm}\usepackage{dcolumn}\usepackage{latexsym}

\makeatother

\usepackage{babel}

\makeatother

\usepackage{babel}

\makeatother

\usepackage{babel}

\begin{document}
\global\long\global\long\global\long\def\etal{\emph{et al.}}

\pacs{03.67.-a, 32.80.Fb, 37.10.Jk, 42.50 Dv}

\preprint{V2}

\title{Highly Efficient State-Selective Submicrosecond Photoionization Detection of Single Atoms}

\author{F.\,Henkel}

\email[]{florian.henkel@physik.uni-muenchen.de}

\author{M.\,Krug}

\author{J.\,Hofmann}

\author{W.\,Rosenfeld}

\author{M.\,Weber}

\affiliation{Fakultät für Physik, Ludwig-Maximilians-Universität München, D-80799
München, Germany}

\author{H.\,Weinfurter}

\affiliation{Fakultät für Physik, Ludwig-Maximilians-Universität München, D-80799
München, Germany}

\affiliation{Max-Planck-Institut für Quantenoptik, D-85748 Garching, Germany}

\date{\today}
\begin{abstract}
We experimentally demonstrate a detection scheme suitable for state
analysis of single optically trapped atoms in less than $1\,\mathrm{\mu s}$
with an overall detection efficiency $\eta$ exceeding $98\%$. The
method is based on hyperfine-state-selective photoionization and subsequent
registration of the correlated photoion-electron pairs by coincidence
counting via two opposing channel electron multipliers. The scheme
enables the calibration of absolute detection efficiencies and might
be a key ingredient for future quantum information applications or
precision spectroscopy of ultracold atoms. 
\end{abstract}
\maketitle
One crucial requirement for quantum computation, quantum communication, and
quantum metrology is the highly efficient measurement of qubit states. For
atomic qubits, the most frequently used fluorescence method allows measuring
with a detection efficiency of almost unity, however, at the cost of comparably
long detection times $(>100\,\mu\mathrm{s})$. In order to reduce the
measurement duration for such systems one can either increase the numerical
aperture of the collection optics \cite{BetterCollectionAngle} or,
alternatively, enhance fluorescence by means of optical cavities
\cite{Cavity}.

Yet, if one intends to apply these approaches for state analysis of many
atoms, e.g. for one-way quantum computation in optical lattices
\cite{Briegel2001}, difficulties arise. Even with fluorescence collection
pushed to the limit \cite{BetterCollectionAngle} the detection times can
hardly be reduced below $10\,\mu\mathrm{s}$ per atom. Alternatively, optical
cavities allow detection times below $1\,\mu\mathrm{s}$ \cite{Terraciano2009}.
However, due to the typically small mode volume and reduced optical access,
scalability to a large number of atoms may be challenging with current state
of the art technologies. Therefore, in order to obtain both scalability and
speed, a completely different approach is required.

\begin{figure}
\includegraphics{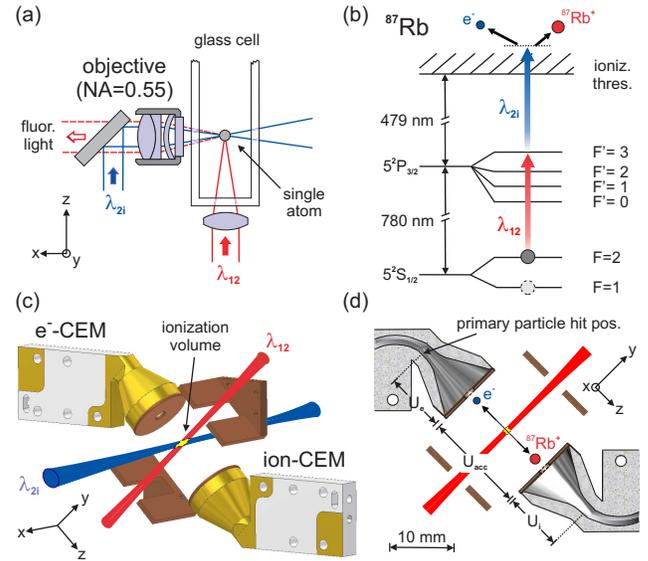} \caption{\label{fig:Skizze}(color
  online). (a) Setup for single atom ionization: Single atoms are trapped in
  an optical dipole trap, prepared into selected hyperfine states and
  subsequently ionized. (b) Photoionization level scheme: Single
  ion-$\mathrm{e}^{-}$ pairs are generated by hyperfine-state-selective,
  resonant two-step, two-color photoionization.  (c) Joint channel electron
  multiplier (CEM) detector: two opposing CEMs and compensation electrodes
  against stray fields are built into a glass cell identical to (a). (d)
  Detector geometry in section view (laser beam waist not to scale).}

\end{figure}

In this Letter, we experimentally demonstrate how hyperfine-state-selective
photoionization and subsequent detection of the generated photoion-electron
pairs enable fast and efficient state analysis. On the example of
single $^{87}\mathrm{Rb}$-atoms, we reach a readout fidelity of $\mathcal{F}=99.2\%$,
an overall detection time below $1\,\mu\mathrm{s}$, and an overall
detection efficiency $\eta$ exceeding $98\%$.

\begin{figure}
\includegraphics{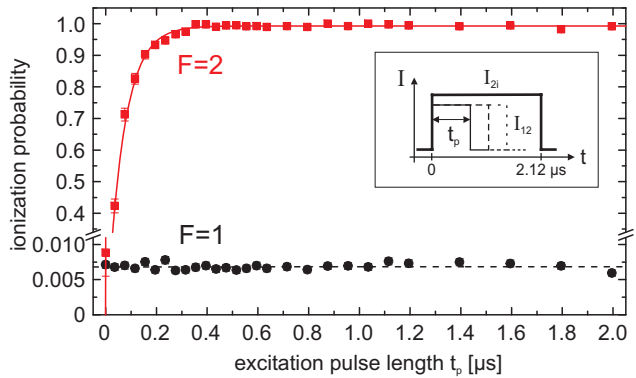}\caption{\label{fig:StateSelectiveIonisation}
(color online). Hyperfine-state-selective, single atom ionization probability
  for different laser pulse lengths.  Atoms in the trap are initially prepared
  either in the $5^{2}S_{1/2}$ $F=2$ ($\Box$) or the $F=1$ ($\circ$) hyperfine
  state, respectively.  Only atoms in $F=2$ driven by both laser fields
  $(\lambda_{12},\lambda_{2\textrm{i}})$ are ionized. Inset: Scheme showing
  the timing of the excitation and ionization pulses.}

\end{figure}

In a first experiment, efficiency, speed and hyperfine-state selectivity of
the photoionization procedure are studied \cite{Zimmermann}. For this purpose,
a single $^{87}\mathrm{Rb}$ atom is loaded into a far off resonance optical
dipole trap in an UHV glass cell setup, similar to \cite{Weber2006}
[Fig.\,\ref{fig:Skizze}(a)]. State selectivity is achieved by a two-step,
two-color photoionization scheme using the $5^{2}P_{3/2},\, F'=3$ level as
resonant, intermediate state [Fig.\,\ref{fig:Skizze}(b),
$\lambda_{12}=780\,\mathrm{nm}$, $\lambda_{2\mathrm{i}}=473\,\mathrm{nm}$].
For that purpose two perpendicular, focused laser beams
$(w_{12}=54\,\mu\mathrm{m}$, $P_{12}=6\,\mu\mathrm{W}$;
$w_{2\textrm{i}}=1.13\,\mu\mathrm{m}$, $P_{2\textrm{i}}=32.8\,\mathrm{mW})$
are overlapped with the optical dipole trap into one common focus
[Fig.\,\ref{fig:Skizze}(a)]. The polarizations of both laser fields are linear
and parallel to the y-axis. During measurements no electric field is applied
and only small magnetic fields $(<50\,\mathrm{mG)}$ are present.

To evaluate the photoionization process, we investigate the ionization
probability of the trapped atom for different excitation pulse lengths
(Fig.\,\ref{fig:StateSelectiveIonisation}). By optical pumping, the atom is
initially prepared either in the $5^{2}S_{1/2},\, F=1$ or $F=2$ hyperfine
ground state, respectively [Fig.\,\ref{fig:Skizze}(b)].  Then, pulses of the
excitation and ionization light are simultaneously applied. While the
ionization pulse length is fixed to $2.12\,\mathrm{\mu s}$ for all
measurements, the excitation pulse length $t_{\textrm{p}}$ is varied between
$36\,\mathrm{ns}$ and $2.0\,\mu\mathrm{s}$
(Fig.\,\ref{fig:StateSelectiveIonisation}, inset). The excitation pulse
operates at a multiple of the saturation intensity of the cycling transition,
yielding the $5^{2}P_{3/2}$ state population approaching one-half after a few
lifetimes of the excited state $(\tau_{\mathrm{exc}}=26.2\,\mathrm{ns})$.

Figure$\,$\ref{fig:StateSelectiveIonisation} shows the ionization
probability after application of a single excitation-ionization sequence.
As the ionization process removes the atom out of the trap, the ionization
probability is derived from atom loss, analyzed by subsequent fluorescence
detection. For atoms initially prepared in $F=2$, the dynamics of
the resonant two-step, two-color photoionization is described by a
rate equation model \cite{Letokhov1979}, valid for $t_{p}$ longer
than the lifetime of the intermediate $5^{2}P_{3/2}$ level. The ionization
probability is \begin{equation}
p_{\textrm{ion,F=2}}(t_{p})=p_{\infty}(1-\exp[-\rho_{\textrm{ee}}\sigma_{2\textrm{i}}\Phi_{2\textrm{i}}t_{p}]),\label{eq:IonisationEfficiency}\end{equation}
 with $p_{\infty}$ being the probability to ionize the atom for $t_{p}\rightarrow\infty$,
$\rho_{\textrm{ee}}\approx\frac{1}{2}$ the steady-state population
of the $5^{2}P_{3/2}$ level, $\sigma_{2\textrm{i}}$ the ionization
cross section \cite{ScatteringCrossSections}, and $\Phi_{2\textrm{i}}$
the photonic flux of the ionizing laser. For the evaluation of $p_{\textrm{ion,F=2}}$,
the parameters $\rho_{\textrm{ee}},\sigma_{2\textrm{i}}$ and $\Phi_{2\textrm{i}}$
are combined into a characteristic $\frac{1}{e}$-ionization time
$\tau=(\rho_{\textrm{ee}}\sigma_{2\textrm{i}}\Phi_{2\textrm{i}})^{-1}$.
To determine $\tau$, a least-square fit according to (\ref{eq:IonisationEfficiency})
is applied for pulse lengths $t_{p}$ up to $475\,\mathrm{ns}$ (red
line, Fig.\,\ref{fig:StateSelectiveIonisation}). With $p_{\infty}=0.993\pm0.001$
deduced by averaging the measured ionization probabilities for $t_{p}>475\,\mathrm{ns}$,
we obtain $\tau=64.4\pm2.8\,\mathrm{ns}$. Thus, after an ionization
time $t_{\textrm{ion}}\equiv6\tau=386\,\mathrm{ns}$ ($F=2$, Fig.\,\ref{fig:StateSelectiveIonisation}),
an ionization probability of $p_{\textrm{ion,F=2}}(t_{\textrm{ion}})=0.9905\pm0.0010$
is achieved. Other loss mechanisms, as, e.g., possible single-color,
multiple-photon ionization processes or heating are considered. For
that purpose each of the beams is switched on separately, showing
negligible ionization probabilities. In order to demonstrate the hyperfine-state
selectivity of the ionization scheme, we compare this result with
measurements for the atom initially prepared in the $5^{2}S_{1/2},\, F=1$
state. Here, we observe a probability $p_{\textrm{ion,F=1}}$ of $0.0068\pm0.0001$
for all excitation pulse lengths $t_{\textrm{p}}$ ($F=1$, Fig.\,\ref{fig:StateSelectiveIonisation}).

From these measurements, a readout fidelity $\mathcal{F=}\frac{1}{2}[p_{\textrm{ion,F=2}}+(1-p_{\textrm{ion,F=1}})]$,
defined as the average probability to correctly identify the hyperfine
state, can be deduced, resulting in a value of $\mathcal{\mathcal{F}}(t_{\textrm{ion}})=99.19\pm0.05\%$
\cite{AppliedElectricField}. Note, this value also includes preparation
errors, atom loss out of the trap, and errors of the fluorescence
detection and since the observed $p_{\textrm{ion,F=1}}$ is indistinguishable
from measurements without any ionizing light the actual readout fidelity
of the ionization process is even closer to unity.

In a second experiment, the efficiency and time required for detecting the
generated photoion-electron pairs, as well as the sensitive volume of the
detection system are investigated. To detect both ionization fragments, this
system consists of two channel electron multipliers (CEMs,
\cite{SjutsHompage}) whose cone entrances are separated by
$d=15.7\,\mathrm{mm}$. To protect the CEM cones against stray light and to
tailor the electric fields inside the cones, the entrances are covered by
copper plates with an open aperture of $2\,\mathrm{mm}$.  Additional electrodes
next to the CEMs compensate for electric stray fields
[Fig.\,\ref{fig:Skizze}(c)]. This system is mounted in a separate UHV glass
cell setup, where in between both detectors at $z=d/2$,
$^{87}\mathrm{Rb}$-atoms from the thermal background vapor are photoionized
within the overlap of two mutually perpendicular laser beams. The overlap
region can be moved in all three dimensions in order to investigate the
spatial dependence of the detection efficiency (the sensitive volume). Uniform
ionization conditions over the scan region are provided by only weakly
focusing the beams $(w_{12}=26\,\mu\mathrm{m}$, $P_{12}=155\,\mu\mathrm{W}$;
$w_{2\textrm{i}}=43\,\mu\mathrm{m}$, $P_{2\textrm{i}}=164\,\mathrm{mW})$. To
separate the oppositely charged photoionization fragments into their
corresponding CEM, the CEMs are held at different electric potentials defining
the total acceleration voltage $U_{acc}$ [Fig.\,\ref{fig:Skizze}(d)] which in
turn determines the kinetic energy $E_{\mathrm{kin}}$ of the fragments at the
cone entrance \cite{ConeEntrance}. The particular arrangement of the CEMs and
the compensation electrodes ensures that no additional electron/ion
optics is required, thus allowing a large solid angle for optical access.

\begin{figure}
\includegraphics{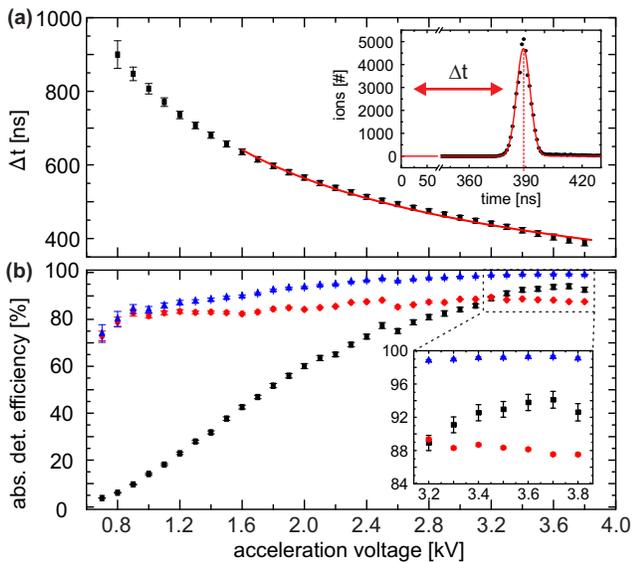}\caption{\label{fig:CEM-Efficiency}(color
  online). (a) Time-of-flight difference $\Delta t$ of generated
  $^{87}\mathrm{Rb}$-ions to their corresponding photoelectrons for different
  acceleration voltages $U_{acc}$. Inset: Sample histogram of time
  differences between $^{87}\mathrm{Rb}$-ion and electron detections for
  $U_{acc}=3.8\,\mathrm{kV}$.  (b) Absolute detection efficiency for
  $^{87}\mathrm{Rb}$-ions ($\Box$, black), electrons ($\circ$, red) for
  different acceleration voltages and the calculated, total detection
  efficiency ($\vartriangle$, blue).  Inset: Zoom for acceleration voltages
  from $3.2$ to $3.8\,\mathrm{kV}$.}
\end{figure}

In order to determine the duration $t_{\textrm{det}}$ of the detection
process, the time from the ionization event to the detection of the
macroscopic pulses at the anodes of the CEMs is investigated
\cite{GeneralCEMgain}.  It is composed of the respective flight times
$(t_{e},t_{i})$ of the two ionization fragments until the primary particle hit
in the corresponding detector [Fig.\,\ref{fig:Skizze}(d)] and the transit time
of the electron avalanche inside the CEM channels. In the experiment only the
time-of-flight difference $\Delta t=t_{i}-t_{e}$ is accessible
[Fig.\,\ref{fig:CEM-Efficiency}(a)]. It is deduced from a Gaussian fit of the
detection time differences [Fig.\,\ref{fig:CEM-Efficiency}(a), inset]. The
temporal spread of the correlation peak [indicated by the error bars in
Fig.\,\ref{fig:CEM-Efficiency}(a)] remains narrow for a wide range of
acceleration voltages $(1.6-3.8\,\mathrm{kV})$ with
$\mathrm{FWHM}\leq8.5\,\mathrm{ns}$. The measured time-of-flight differences
can be modeled assuming acceleration of the ion or electron in a homogeneous
field $E_{\textrm{acc}}$ up to the CEM entrance and further
deceleration (acceleration) within the respective CEM. The model holds for
$U_{acc}$ above $1.6\,\mathrm{kV}$ [red line, Fig.\ref{fig:CEM-Efficiency}(a)],
below this value the actual field configurations between and inside the CEMs
have to be taken into account in more detail. At $U_{acc}=3.8\,\mathrm{kV}$,
we observe a time-of-flight difference of $\Delta
t=388.81\pm0.01\,\mathrm{ns}$.  According to the model we obtain a
photoelectron flight time of $t_{e}=0.95\,\mathrm{ns}$ for this acceleration
voltage, which together with the CEM transit time of $26\,\mathrm{ns}$
\cite{SjutsHompage} sums up to a detection time of $t_{\textrm{det}}=\Delta
t+t_{e}+t_{\mathrm{transit}}=415.8\,\mathrm{ns}$.

For many applications the crucial parameter is a high detection efficiency
\cite{Campey2006,Rosenfeld2009}, i.e. in our case the total efficiency
of collecting the respective fragment into the CEM and converting
it into an observable electron avalanche. In order to optimize the
efficiency for our detection system, the positions of the primary
particle hit inside the CEMs are designed to be under grazing incidence
at the channel walls [Fig.\,\ref{fig:Skizze}(d)] and not in the
CEM cones \cite{Fricke1980,Gilmore1999}. Absolute detection efficiencies
can be determined by coincident detection of the oppositely charged
ionization fragments with two particle detectors \cite{2D-scan-efficiency}.
Such a calibration is in perfect conceptual correspondence to $4\pi\beta\gamma$-coincidence
counting \cite{Dunworth1940} or absolute photodetector calibration
via photon pairs \cite{Kwiat1994}. Accordingly, the CEM ion and $\mathrm{e}^{-}$-detection
efficiencies $\eta_{i},\eta_{e}$ are given by
\begin{equation}
\eta_{i}=\frac{N_{c}}{N'_{e}},\ \ \eta_{e}=\frac{N_{c}}{N'_{i}},\label{eq:CEMDetectorEfficiency}\end{equation}
where $N'_{i}=N_{i}-N_{bi}$ and $N'_{e}=N_{e}-N_{be}$ are the background
corrected ion and e$^{-}$ single counts and $N_{c}$ is the number of
coincidences. The latter is obtained from the detection time difference
histograms [Fig.\,\ref{fig:CEM-Efficiency}(a), inset] with a coincidence
time window starting $20\,\mathrm{ns}$ before the correlation peak
and ending $80\,\mathrm{ns}$ after it. This choice results from the
presence of a few late ion detections. Accidental coincidences can
be neglected, as the fraction of accidental to true coincidences is
smaller than $10^{-4}$. The background events $N_{bi},N_{be}$ are
measured with the excitation laser turned off, leaving only the ionization
laser switched on. Figure\,\ref{fig:CEM-Efficiency}(b) shows the
absolute detection efficiencies for ions and electrons for different
acceleration voltages $U_{acc}$. For values up to $3.8\,\mathrm{kV}$, the ion
detection efficiency $\eta_{i}$ increases in qualitative agreement
with previous studies for different ion species \cite{Burrows1967,Gilmore1999}.
In contrast to this, the measured electron detection efficiency $\eta_{e}$
remains almost constant. At $U_{acc}=3.8\,\mathrm{kV}$,
for a measurement time of $60\,\mathrm{s}$ we observe $N_{i}=53762$,
$N_{bi}=2235$, $N_{e}=196547$, $N_{be}=147845$ and $N_{c}=45099$
(including ion and electron dark counts $N_{di}\sim2100$ and $N_{de}\sim15000$,
respectively), resulting in an absolute CEM detection efficiency of
$\eta_{i}=0.926\pm0.010$ and $\eta_{e}=0.875\pm0.002$ \cite{Error}.
From these efficiencies, a total efficiency for the detection of a
photoionized neutral atom can be determined. As it is sufficient to
detect the photoion \emph{or} the photoelectron, the total detection
efficiency $\eta_{\textrm{det}}$ is \begin{equation}
\eta_{\textrm{det}}=1-(1-\eta_{i})(1-\eta_{e})=\eta_{e}+\eta_{i}-\eta_{i}\eta_{e}.\label{eq:CEMAbsoluteDetectionEfficiency}\end{equation}
 Using the above values at $3.8\,\mathrm{kV}$, a total detection
efficiency of $\eta_{\textrm{det}}=0.991\pm0.002$ is derived.

\begin{figure}
\includegraphics{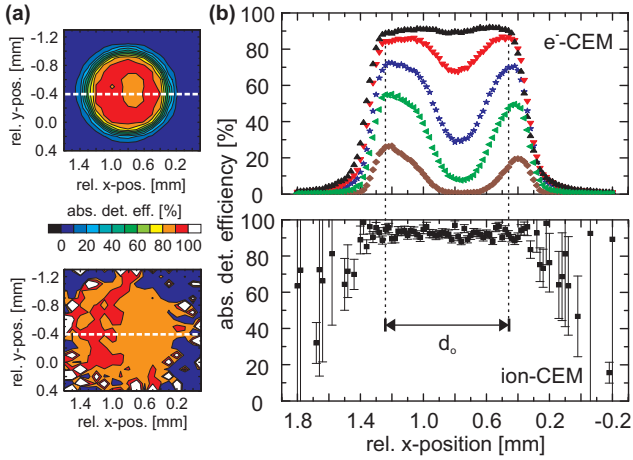}\caption{\label{fig:CEM-Efficiency-Area}(color
  online). (a) 2D scans of the sensitive areas at a gain voltage of
  $2.8\,\mathrm{kV}$ for the $\mathrm{e}^{-}$ CEM (top) and ion CEM (bottom)
  for an acceleration voltage of $3.8\,\mathrm{kV}$.  Dashed lines indicate
  respective line scans in (b). (b) Line scans through the center at
  $y=-0.4\,\mathrm{mm}$. Electron detection efficiencies (top) at different
  gain voltages ($\Diamond=2.4\,\mathrm{kV}$ (brown);
  $\vartriangleleft=2.5\,\mathrm{kV}$ (green); $\star=2.6\,\mathrm{kV}$
  (blue); $\triangledown=2.8\,\mathrm{kV}$ (red);
  $\vartriangle=3.2\,\mathrm{kV}$ (black)). $^{87}\mathrm{Rb}$-ion detection
  efficiency (bottom) for a gain voltage of $2.7\,\mathrm{kV}$
  ($\Box$). The acceleration voltage is the same as in (a).}
\end{figure}

Finally, we analyze the spatial dependence of the detection efficiency, i.e.,
the sensitive volume of the detection
system. Figure\,\ref{fig:CEM-Efficiency-Area}(a) shows the spatial
distribution of the efficiency determined by a 2D scan in the x-y plane at
$d/2$ in between both CEMs. Here, for both detectors, a roughly circular
distribution is observed. A scan along the x direction through the center of
the area is depicted in Fig.\,\ref{fig:CEM-Efficiency-Area}(b) for different
CEM gain voltages at $U_{acc}=3.8\,\mathrm{kV}$. Uniform efficiencies over the
full area [see $\mathrm{e}^{-}$-CEM, Fig.\,\ref{fig:CEM-Efficiency-Area}(b)]
are achieved for high gain voltages \cite{OldDetectors}, similar to previous
experiments with electron and ion beams \cite{Gilmore1999}.  From these
measurements we obtain a sensitive area with a diameter of
$d_{\circ}=0.84\,\mathrm{mm}$ where $\eta_{\textrm{det}}$ exceeds
$98.8\,\%$. Corresponding measurements at different z positions display a
similar spatial behavior, indicating that the sensitive volume has a
longitudinal extension of at least $5\,\mathrm{mm}$. By moving the ionization
region closer to the ion-CEM and adapting $U_{acc}$, even shorter ion
flight times can be achieved, retaining high detection efficiencies.

Combining all measurement results, an overall efficiency
$\eta=p_{\textrm{ion,F=2}}\eta_{\textrm{det}}$ for state-selective detection
of a single, neutral atom stored in an optical dipole trap and its
corresponding detection time
$t_{\textrm{tot}}=t_{\textrm{ion}}+t_{\textrm{det}}$ can be estimated. At an
acceleration voltage of $U_{acc}=3.8\,\mathrm{kV}$, an overall
detection efficiency of $\eta=0.982\pm0.002$ with an overall detection time
$t_{\textrm{tot}}=802\pm17\,\mathrm{ns}$ is determined.

In this Letter, we have shown a highly efficient, state-selective and
fast ionization detection system for optically trapped neutral atoms
which allows deterministic state analysis at the single-atom
level. Though the detection process removes the analyzed atom from
the trap, the system has a number of possible applications, for example
combined with single atom traps it allows to determine partial and
total photoionization cross sections free from ensemble averages \cite{ScatteringCrossSections}
or together with coherent stimulated Raman adiabatic passage techniques
the state analysis of Zeeman qubits \cite{Rosenfeld2009}. Contrary
to other approaches for single-atom state analysis, the geometry of
the detection system provides a comparably large sensitive volume
and high optical access. In the future it might thus be applied, e.g.,
for imaging \cite{Weiss2007} and site-specific readout of atoms in
optical lattices where pulsed, scanning excitation schemes operating
at high powers can enable readout times below $100\,\mathrm{ns}$
per atom. Furthermore, it could also be used for \textit{in situ}, real-time
probing of ultracold ensembles with sub-poissonian accuracy \cite{Campey2006}
or as detector for a loophole-free test of Bell's inequality with
a pair of trapped atoms at remote locations \cite{Rosenfeld2009}. 
\begin{acknowledgments}
We acknowledge stimulating discussions with T. W.\,Hänsch, J.\,Volz,
T.\,Schätz, H.\,Sjuts, and craftsmanship of the LMU workshop. Financial
support was provided by the EU, the Elite Network of Bavaria, and the DFG
through the projects Q-ESSENCE, QCCC, and MAP, respectively.
\end{acknowledgments}


\begin{thebibliography}{23}
\bibitem{BetterCollectionAngle} Y. R. P. Sortais \etal, Phys. Rev.
A \textbf{75}, 013406 (2007); M. K. Tey \etal, Nature Physics \textbf{4},
924 (2008); R. Maiwald \etal, Nature Physics \textbf{5}, 551 (2009).

\bibitem{Cavity} I. Teper, Y.J. Lin, and V. Vuletic, Phys. Rev. Lett.
\textbf{97}, 023002 (2006); J. Bochmann \etal, Phys. Rev. Lett. \textbf{104},
203601 (2010); R. Gehr \etal, Phys. Rev. Lett. \textbf{104}, 203602
(2010).

\bibitem{Briegel2001} R. Raussendorf, and H. J. Briegel, Phys. Rev.
Lett. \textbf{86}, 5188 (2001); O. Mandel \etal, Nature \textbf{425},
937 (2003).

\bibitem{Terraciano2009}M. L. Terraciano et al., Nature Physics\textbf{
5}, 480 (2009).

\bibitem{Zimmermann}A. Stibor et al. Phys. Rev. A \textbf{76}, 033614
(2007); A. Günther et al. Phys. Rev. A \textbf{80}, 011604(R) (2009).

\bibitem{Weber2006} M. Weber \etal, Phys. Rev. A \textbf{73}, 043406-7
(2006).

\bibitem{Letokhov1979} V. S. Letokhov, V. I. Mishin, and A. A. Puretzky,
Prog. Quantum Electron. \textbf{5}, 139 (1979).

\bibitem{ScatteringCrossSections} T. Dinneen \etal, Opt. Lett. \textbf{17},
1706 (1992); C. Gabbanini, S. Gozzini, and A. Lucchesini, Opt. Comm.
\textbf{141}, 25 (1997); D. Ciampini \etal, Phys. Rev. A \textbf{66},
043409 (2002).

\bibitem{AppliedElectricField} For later applied acceleration voltages of
  $3.8\,\mathrm{kV}$, the Stark-shifts of the Zeeman-sublevels of the
  $5^{2}S{}_{1/2}$ and $5^{2}P_{3/2}$ states are more than order of
  magnitude smaller than the natural linewidth of the transition. Thus, the
  spin-selectivity of the ionization process should not be affected.

\bibitem{SjutsHompage} Type KBL10RS45-V2, Dr. Sjuts Optotechnik GmbH.

\bibitem{ConeEntrance} The compensation electrodes modify the electric
  potential at the ioization region ($z=d/2$) such, that ions experience
  $71\%$ of the applied acceleration voltage $U_{acc}$. The remaining $29\%$
  of $U_{acc}$ determine the kinetic energy of the electrons. With the current
  setup, the maximum applied acceleration voltage is $3.8\,\mathrm{kV}$.

\bibitem{GeneralCEMgain}If not specified otherwise the CEM gain voltage
is $3.0\,\mathrm{kV}.$

\bibitem{Rosenfeld2009} J.Volz \etal, Phys. Rev. Lett. \textbf{96},
030404 (2006); W. Rosenfeld \etal, Adv. Sci. Lett. \textbf{2}, 469
(2009).

\bibitem{Campey2006} T. Campey \etal, Phys. Rev. A \textbf{74},
043612-9 (2006).

\bibitem{Fricke1980} J. Fricke, A. Müller, and E. Salzborn, Nucl.
Inst. Meth. \textbf{175}, 379 (1980).

\bibitem{Gilmore1999} M. P. Seah, and G. C. Smith, Rev. Sci. Instrum.
\textbf{62}, 62 (1991); I. S. Gilmore, and M. P. Seah, Appl. Sur.
Sci. \textbf{144}, 113 (1999).

\bibitem{2D-scan-efficiency} E.S. Fry, T. Walther, and S. Li, Phys.
Rev. A \textbf{52}, 4381 (1995); B. Brehm \etal, Meas. Sci. Tech.
\textbf{6}, 953 (1995).

\bibitem{Dunworth1940} J. V. Dunworth, Rev. Sci. Instrum. \textbf{11},
167 (1940).

\bibitem{Kwiat1994} P. Kwiat \etal, Appl. Opt. \textbf{33}, 1844
(1994).

\bibitem{Burrows1967} C. N. Burrows, A. J. Lieber, and V. T. Zaviantseff,
Rev. Sci. Instrum. \textbf{38}, 1477 (1967); A. Egidi \etal, Rev.
Sci. Instrum. \textbf{40}, 88 (1969); B. Tatry, J. M. Bosqued, and
H. Reme, Nucl. Inst. Meth. \textbf{69}, 254 (1969); J. Fox, R. L.
Fitzwilson, and E. W. Thomas, J. Phys. E \textbf{3}, 36 (1970).

\bibitem{Error}The statistical errors of the detection efficiencies
[error bars in Fig.\ref{fig:CEM-Efficiency}(b)] are given by the
variances $\sigma_{\eta_{i,e}}^{2}=(N_{c}/N_{e,i}^{'})^{2}\,(1/N_{c}-1/N'_{e,i}+2N_{be,bi}/N_{e,i}^{'2})$.

\bibitem{OldDetectors} The high gain voltages result from detector aging.

\bibitem{Weiss2007} K. D. Nelson, X. Li, and D. S. Weiss, Nature
Physics \textbf{3}, 556 (2007); T. Gericke \etal, Nature Physics
\textbf{4}, 949 (2008); M. Karski \etal, Phys. Rev. Lett. \textbf{102},
053001 (2009); W. S. Bakr \etal, Nature \textbf{462}, 74, (2009);
J. Sherson \etal, Nature \textbf{467}, 68, (2010). 
\end{thebibliography}
\end{document}